# Sub-DM: Subspace Diffusion Model with Orthogonal Decomposition for MRI Reconstruction

Yu Guan, Qinrong Cai, Wei Li, Qiuyun Fan, Dong Liang, *Senior Member, IEEE*, and Qiegen Liu, *Senior Member, IEEE*

*Abstract*—Diffusion model-based approaches recently achieved remarkable success in MRI reconstruction, but integration into clinical routine remains challenging due to its time-consuming convergence. This phenomenon is particularly notable when directly apply conventional diffusion process to k-space data without considering the inherent properties of k-space sampling, limiting k-space learning efficiency and image reconstruction quality. To tackle these challenges, we introduce subspace diffusion model with orthogonal decomposition, a method (referred to as Sub-DM) that restrict the diffusion process via projections onto subspace as the k-space data distribution evolves toward noise. Particularly, the subspace diffusion model circumvents the inference challenges posed by the complex and high-dimensional characteristics of k-space data, so the highly compact subspace ensures that diffusion process requires only a few simple iterations to produce accurate prior information. Furthermore, the orthogonal decomposition strategy based on wavelet transform hinders the information loss during the migration of the vanilla diffusion process to the subspace. Considering the strategy is approximately reversible, such that the entire process can be reversed. As a result, it allows the diffusion processes in different spaces to refine models through a mutual feedback mechanism, enabling the learning of accurate prior even when dealing with complex k-space data. Comprehensive experiments on different datasets clearly demonstrate that the superiority of Sub-DM against state of-the-art methods in terms of reconstruction speed and quality.

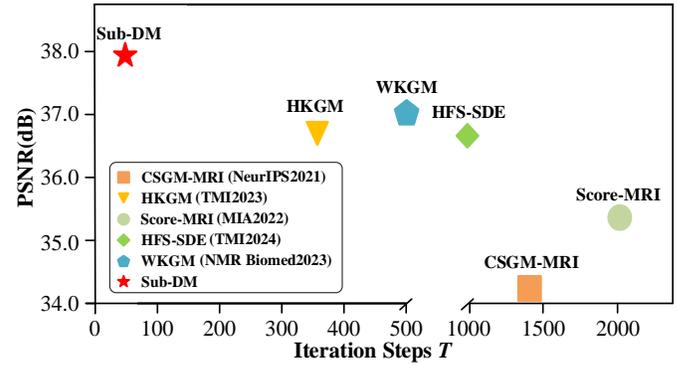

**Fig. 1.** Convergence analysis of CSGM-MRI, HKGM, Score-MRI, HFS-SDE, WKGM, and Sub-DM in terms of PSNR versus the iteration steps for brain image reconstruction at $R$=8 under Poisson sampling.

## I. INTRODUCTION

Magnetic Resonance Imaging (MRI) is one of the most widely used imaging modalities due to its excellent soft tissue contrast, but it has prolonged and costly scan sessions [1], [2]. This limitation has galvanized innovations to accelerate the MRI process, all with the common goal of drastically reducing scan time without compromising image quality. A fundamental solution is to shorten scan times by under-sampled k-space data and solve an ill-posed inverse problem to reconstruct images. From sparsity-driven compressed sensing [3]-[5] to deep-learning-based model [6]-[8], significant progress has been made in MRI reconstruction. However, many existing methods are limited by suboptimal capture of the data distribution and reliance on fully-sampled acquisitions for model training.

The desire for more robust and efficient techniques in MRI reconstruction has led to the development of pioneering approaches, among which diffusion models (DMs) [9]-[12] have recently shown to be promising. As a promising surrogate, DMs provide a more accurate representation of the data distribution and shows great potential to reconstruct MR images. One of the notable advancements in this area was initially proposed by Jalal *et al*. [13], where they trained the DMs using Langevin dynamics without making any assumptions on the measurement system, yielding competitive reconstruction results for both in-distribution and out-of-distribution data. Inspired by this innovative work, Song *et al*. [14] were committed to expanding the theoretical framework of DMs, who successfully extended them to medical image reconstruction tasks. Furthermore, Chung *et al*. [15] demonstrated that score-based diffusion models trained solely on magnitude images can be utilized for reconstructing complex-valued data. Luo *et al*. [16] described a comprehensive approach using data-driven Markov chains for MRI reconstruction which not only facilitates efficient image reconstruction across variable sampling schemes, but also enables the generation of uncertainty maps.

Despite demonstrating high-quality sampling in MRI reconstruction, DMs suffer from slow sampling processes and high

This work was supported in part by the National Key Research and Development Program of China under Grant 2023YFF1204300 and Grant 2023YFF1204302, in part by the National Natural Science Foundation of China under Grant 62122033. (Y. Guan and Q. Cai are co-first authors) (Corresponding authors: D. Liang and Q. Liu).

Y. Guan is with the School of Mathematics and Computer Sciences, Nanchang University, Nanchang 330031, China. (guanyu@email.ncu.edu.cn).

Q. Fan is with the Academy of Medical Engineering and Translational Medicine, Medical School, Faculty of Medicine, Tianjin University, Tianjin 300072, China. (fanqiuyun@tju.edu.cn).

D. Liang is with the Lauterbur Research Center for Biomedical Imaging and the Research Center for Medical AI, Shenzhen Institute of Advanced Technology, Chinese Academy of Sciences, Shenzhen 518055, China. (dong.liang@siat.ac.cn).

Q. Cai, W. Li, and Q. Liu are with the School of Information Engineering, Nanchang University, Nanchang 330031, China. ({caiqinrong, 416100230053}@email.ncu.edu.cn, shaoyuwang22@gmail.com, liuqiegen@ncu.edu.cn)).

computational burden due to the need for hundreds of reverse steps for image generation [17], [18]. To overcome the inherent drawbacks caused by DMs, researchers turned the focus to exploring optimization techniques for DMs to decrease sampling time. Peng *et al*. [19] proposed a representative work that considered rescaling the diffusion step size during inference to accelerate image sampling, but this can potentially reduce the accuracy of reverse diffusion steps. Instead of directly altering the network structure to minimize iteration time, recent innovative studies have reengineered the diffusion process to unfold in the latent space, consequently reducing memory consumption. A straightforward extension of the latent space method proposed by Li *et al*. [20], which considered compressing the latent features in the DM into a low-dimensional latent space to reduce computational complexity and the number of iteration steps. While the potential is undeniably promising, these methodologies concentrate on the model optimization while neglecting the intricate characteristics of k-space data, thereby inadvertently increasing the difficulty of achieving fast convergence.

Reevaluating the essence of MRI reconstruction reveals that the main difference between MR imaging and other medical imaging modalities is the control over the k-space data acquisition and how it can be managed to yield an adequate reconstructed image [21], [22]. To effectively optimize the sampling process of the DMs for this task, a comprehensive analysis of the characteristics of k-space data is essential. Peng *et al*. [23] have considered this by conducting Hankel transformations on the k-space data and extracting relevant blocks for the phase of training. This approach enables them to generate samples from a training set that can be as minimal as one k-space data. Moreover, an optimized subspace formed by frequency separation of k-space data is proposed by Cao *et al*. [24], which ensured determinism in the fully-sampled low-frequency regions and accelerated the sampling procedure of reverse diffusion. The success of above methods stems from their ability to retain valid information in k-space data while diminishing its complexity. Hence, it is crucial to prevent the diffusion process from occurring in the space of complete data, as this can lead to refined training coverage and improved performance.

Drawing inspiration from this methodology, this work proposed a pioneering **Sub**space **D**iffusion **M**odel for MRI reconstruction (**Sub-DM**) that integrates orthogonal decomposition for dimensionality modification. One of the innovative aspects of Sub-DM is its emphasis on the role of adjusting the diffusion process to optimize model inference runtimes. Specifically, we employ orthogonal decomposition to extract feature from high-dimensional k-space data as the noise perturbations increase, thereby migrating the diffusion process to a lower-dimensional subspace and avoiding continuous sampling in the intricate full-space. This strategy ultimately enables the sampling process to be implemented within a few diffusion steps and significantly enhances the speed of image reconstruction, as shown in Fig. 1. Moreover, to prevent the loss of information during the diffusion process across different spaces, we employ orthogonal wavelet transforms, capitalizing on their invertibility, as the decomposition operator for dimensionality reduction of k-space data. In this way, it not only reduces computational complexity but also ensures effective learning during the migration process, thereby enabling rapid and accurate MRI reconstruction. The main contributions of this work are summarized as follows:

- To bridge the existing gap between sampling acceleration and generation quality in DMs, we adopt a different perspective to accelerate the diffusion process by reducing the dimensionality of the original signal through orthogonal decomposition, thereby improving the efficiency and obtaining better reconstruction performance.
- Analyzing the distinctive attributes of k-space data, we employ wavelet transforms to decompose it into multiple orthogonal components. This strategy effectively extracts relevant features and concurrently optimizes the solution space for the reverse diffusion process.
- Comprehensive experiments on different datasets demonstrate that Sub-DM achieves faster convergence speed and preserves high reconstruction accuracy under highly under-sampling rates (i.e., $10\times$, $12\times$).

The following sections of the paper are organized as follows: Section II provides a brief overview of related works. Section III presents the core concept of the proposed method. Section IV details the experimental settings and results. Section V offers a succinct discussion and Section VI concludes the work.

## II. RELATED WORK

### A. Problem Formulation

MRI reconstruction is a challenging inverse problem due to the under-sampling operation in k-space. Its objective is to recover the original k-space signal $k \in \mathbb{C}^d$ from the complex-valued measurement $f \in \mathbb{C}^d$. Mathematically, the forward model of this task can be expressed using the following formulation:

$$f = Ak + \eta, \qquad (1)$$

where $A = P\mathbb{F}S$ is the imaging operator that captures the influence of the k-space under-sampling pattern $P$ and coil sensitivity maps $S$, $\mathbb{F}$ denotes the Fourier transform matrix and $\eta$ represents the measurement noise.

Due to the imaging operator $A$ is often rank-deficient, making the recovery of $k$ in Eq. (1) an ill-posed problem. Hence, a regularization prior $R(k)$ is routinely incorporated into the data consistency term to constrain the solution space as follows:

$$k^* = \min_k \{\|Ak - f\|_2^2 + \lambda R(k)\}, \qquad (2)$$

where $k^*$ is the reconstruction and $\lambda$ is a prior knowledge-guided regularization parameter. A growing body of work demonstrates that DMs have recently become the predominant tools for prior extraction. In this work, we aim to optimize the representation of k-space data to enhance its suitability for model training, thereby improving the DM's convergence rate and reconstruction efficiency.

### B. Strategy for Optimizing K-space Data

Considering the distinct underlying information between k-space data and other medical imaging data, the optimization strategies used in k-space data plays an important role in MRI reconstruction [25], [26]. Such strategies reduce the complexity of k-space data through preprocessing, thereby facilitating better integration with model structures. A growing body of work indicate that they are suitable for improving the performance of MRI reconstruction and are flexible in adapting to various models. One most typically employed procedure involves constraining k-space data using artificial masks. For example, Xie *et al*.

[27] applied diffusion process in k-space domain with conditioned under-sampled mask and obtained the high-frequency region of k-space data through the imaging operator. Experimental results shown that they accelerated sampling procedure and outperformed baseline methods.

In contrast to direct truncation methods for handling k-space data, an alternative optimization strategy employing weighted techniques offers a smoother approach [28], [29]. The main motive of the weighted function is to get the uniform distribution of spatial frequency in k-space data that are usually not uniform. Specifically, the uniform distribution of spatial frequency in k-space data can be obtained by multiplying weighted function to the partial k-space. Following this idea to improve the reconstruction accuracy of MRI has been further demonstrated in our earlier work [30], which applied k-space weight-based techniques to capture high-frequency priors. Interestingly, these studies reached similar conclusions while analyzing different reconstruction frameworks. In other words, optimizing k-space data retains its essential information while reducing complexity, thereby enhancing the propagation of features within the network and promoting the data utilization efficiency.

## III. METHOD

### A. Motivation

Unlike other generative models, the distinctive feature of DMs lies in their generative principle, which learns distributions through a more circuitous manner. Specifically, they first define a forward diffusion stage where the input data is gradually perturbed by adding noise, and then learns to reverse the diffusion process to retrieve the desired noise-free data from noisy data samples. Fig. 2 visualizes the two distinct stages of data transfer to noise and generation of new samples in the diffusion process. High image quality has typically been reported with diffusion-based MRI reconstruction. Nevertheless, the above pipeline does not change the dimension of the data throughout the entire diffusion process [11], [31]. It thus requires the reverse process to map a high-dimensional input to a high-dimensional output at every single step, causing heavy computation overheads. Given such a fact, acceleration techniques have recently been considered to speed up the characteristically slow sampling process in regular diffusion models.

A novel strategy rethought the role of Gaussian noise and generalizes the diffusion process using different kinds of degradation approaches. Huang *et al.* [32] proposed a methodology that transitions the degradation operation from the Gaussian white noise to under-sampled k-space data, also designing strategies for starting points and data consistency conditioning to effectively guide and accelerate the reverse process. In contrast to initiate sampling based on under-sampled k-space data, Qiao *et al.* [33] constructed an alternative degradation operation that obtains preliminary guidance images through a separate reconstruction method. While promising, this approach involves implementation of a second reconstruction procedure. Another powerful technology is to train DMs with small step size and to rescale to large step sizes during inference. Gungor *et al.* [34] considered rescaling the diffusion step size via adversarial mapping over large reverse diffusion steps to accelerate image sampling, but reverse diffusion steps can potentially have suboptimal accuracy. Besides varying diffusion in image space, a group of competitive methods [35], [36] directly apply diffusion process in a low dimensional latent space which obtained by an autoencoder. However, these methods significantly modify the original formulation of DMs, such that exact likelihood evaluation and controllable generation become considerably more difficult.

In addition, most approaches overlooking the latent characteristics of k-space data compared to natural images. The oversight significantly diminishes the performance of models so that DMs with strong generative capabilities struggle to learn the distribution of k-space data. One of the primary reasons is that the diffusion process must learn the distribution of high-dimensional latent variables over the entire space, even in areas very far from the data manifold [37]. Due to the curse of dimensionality, much of this space may never be accessed during training, and the accuracy of the model in these regions is called into question by the uncertain extrapolation abilities of networks [38]. Therefore, learning to match a lower-dimensional distribution may lead to refined training coverage and further improved performance.

To address these challenges from a dimensionality perspective, subspace is constructed during the diffusion migration process that focus solely on the essential information within k-space. The essence of this migration mechanism is to reduce the dimensionality of the k-space data, so that the distribution can be learned more accurately even in the case of high noise levels. At the same time, the introduction of orthogonal decomposition technology effectively avoids the loss of information during the migration process. Based on such a decomposition, we theoretically ensure arbitrary transitions across different diffusion space and the smooth progression of the reverse process, thereby enhancing the convergence of sampling. A sketch of the proposed algorithm is shown in Fig. 3.

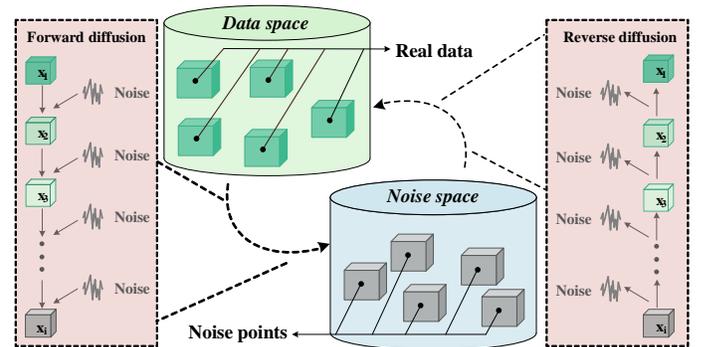

**Fig. 2.** Visual schematic of diffusion process for DMs. Traditional DMs progressively blend noise points with the real data across sequential steps until it evolves into a noise distribution, after which a reverse process is applied at each sampling procedure to neutralize the noise incrementally.

### B. Migrating Subspace to Diffusion Process

*Diffusion Process:* Suppose we are given a k-space MR dataset $k$ in the score-based diffusion model, where each datapoint is independently drawn from an underlying data distribution $p(k)$. With a sequence of positive noise scales $\sigma_{min} = \sigma_0 < \cdots < \sigma_t < \cdots \sigma_T = \sigma_{max}$, the forward diffusion process for $t = 0, \cdots, T$ is defined by an Ito stochastic differential equation (SDE) which iteratively adds the Gaussian white noise and progressively maps the distribution of data $p(k)$ into the normal distribution. In a general case, it is associated with a noising process $k_t$, with $k_0$ distributed according to the data distribution and satisfying:

$$dk_t = \mathbf{f}(k,t)dt + \mathbf{g}(t)du_t. \quad (3)$$

A concrete formulation of a score-based diffusion model requires a choice of forward diffusion process, specified by $\mathbf{f}(k,t)$ and $\mathbf{g}(t)$. Almost always, these are chosen to be isotropic of the form:

$$\mathbf{f}(k,t) = f(t)k, \quad \mathbf{g}(t) = g(t)\mathbf{I}_d, \quad (4)$$

where $\mathbf{I}_d$ is an identity matrix and $d$ is the data dimensionality. For example, the variance exploding (VE) SDE has $f(t) = 0$ and $g(t) = \sqrt{d[\sigma^2(t)]/dt}$ so that it can be parameterized as:

$$dk_t = \sqrt{d[\sigma^2(t)]/dt}\, \mathbf{I}_d du_t, \quad (5)$$

where $\sigma^2(t)$ is the variance of the perturbation kernel at time $t$.

Accordingly, the target of score-based diffusion model is to learn how to reverse the chain and restore the data distribution $p(k)$. Under mild assumptions, for any $T \geq 0$, the reverse diffusion process satisfies:

$$dk_t = \left[\frac{d[\sigma^2(t)]}{dt} \nabla_k log p_{\sigma_t}(k)\right] dt + \sqrt{\frac{d[\sigma^2(t)]}{dt}} d\bar{u}_t. \quad (6)$$

Once an accurate approximation of the Stein score $\nabla_k log p_{\sigma_t}(k)$ is estimated for all $t$, we can derive the reverse diffusion process and simulate it to sample from $p(k)$.

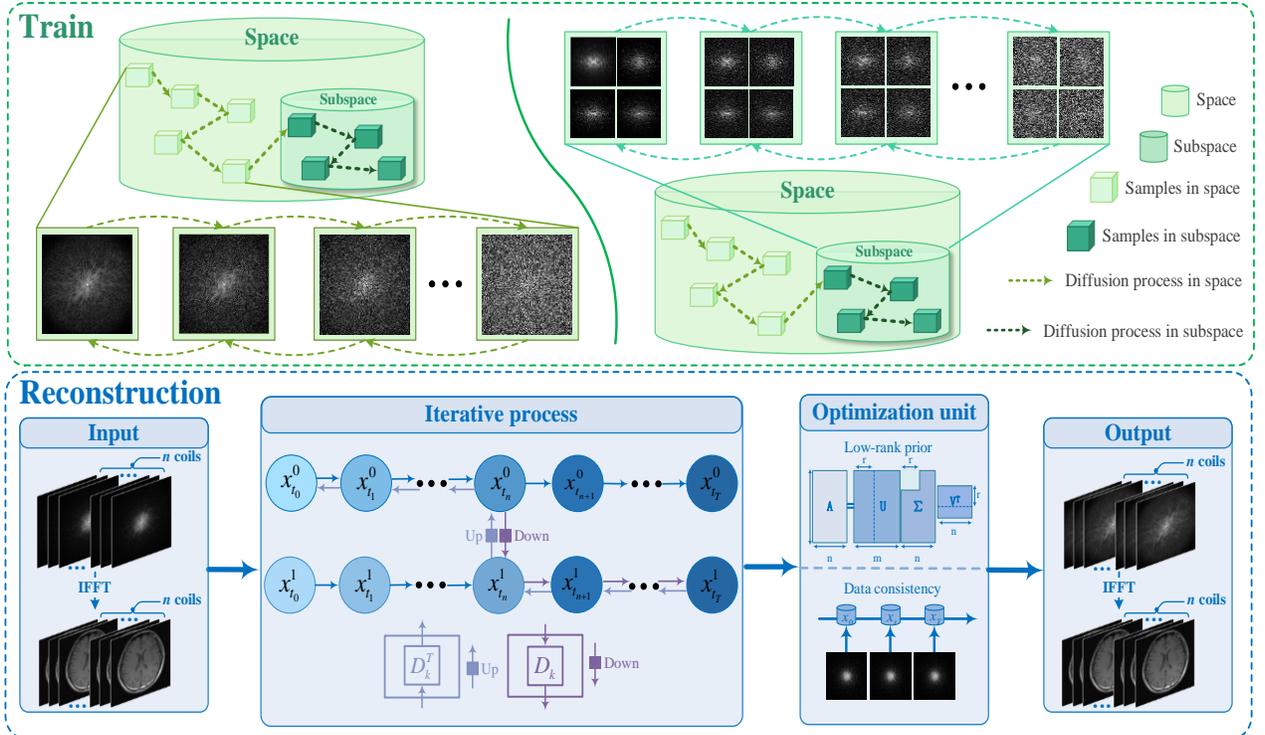

**Fig. 3.** Architecture of Sub-DM based on subspace learning. In the training phase, k-space data undergoes diffusion transformations across two distinct spaces. The original k-space data diffuses in the full space, while the orthogonally decomposed k-space data components diffuse in the subspace. During the reconstruction phase, the dimension of the under-sampled k-space data is dynamically changed by orthogonal decomposition and iteratively reconstructed in various diffusion spaces. Upon completion of the iterations, an optimization module is integrated to enhance the sampling quality.

*Subspace Diffusion:* Considering that in practical MR imaging applications, the target k-space data typically resides near the linear sub-region, such that under isotropic forward diffusion, the components of the data orthogonal to the space become Gaussian significantly compared to the general components. For the reason, we thus propose that at a certain point in time, the diffusion process is migrated to an isotropic subspace $\mathbb{S}$. Specifically, the forward diffusion begins in the full space, but is projected and restricted to smaller subspace as time goes on. For any diffusion process with a form similar to the Eq. (4), we define the corresponding subspace diffusion as follows. Divide the diffusion time $(0, T)$ into individual subintervals with $m$ as the time point, i.e., $(t_0, \cdots, t_m), (t_{m+1} \cdots, t_T)$. Subsequently, diffusion process in subspace can be redefined as follows for each interval $t_0 < t_m < t_T$:

$$\mathbf{g}(t) = g(t)\mathbf{Q}_m\mathbf{Q}_m^T, \quad (7)$$

where $\mathbf{Q}_m \in \mathbb{C}^{d \times d}$ is the orthogonal matrix and its orthonormal columns vectors span the subspace $\mathbb{S}$, $\mathbf{Q}_m^T$ is the transposed matrix which can satisfy $\mathbf{Q}_m\mathbf{Q}_m^T = \mathbf{I}_d$.

Mathematically, these definitions clarify that diffusion process is coupled or constrained to occur in smaller subspaces defined by $\mathbf{Q}_m$ in the interval $(t_m, t_T)$. Turning to drift coefficient $\mathbf{f}(k, t)$, it can be expressed as:

$$\mathbf{f}(k,t) = f(t)k + \sum_{m+1}^{T} \delta(t-t_m)(\mathbf{Q}_m\mathbf{Q}_m^T - \mathbf{I}_d)k, \quad (8)$$

where $\delta(\cdot)$ is the Dirac delta. Eq. (8) states that at time $t_m$, $k$ is projected onto the subspace. Fig. 3 illustrates the high-level concept of subspace diffusion, along with additional properties elaborated in Appendix A. On this basis, we migrate the diffusion process into the subspace, which decreases the dimensionality as time evolves. Learning to match a lower-dimensional score function may lead to refined training coverage and further improved performance.

*Orthogonal Decomposition:* To limit the loss of the information during the dimensionality reduction process, we employ orthogonal decomposition to decrease the dimensionality of the k-space data. Specifically, we concatenate diffusion processes with different dimensions into an entire Markov chain by orthogonal decomposition, while we elaborately design each process so that the information loss induced by orthogonal decomposition is negligible. Meanwhile, the control on information loss ensures that the orthogonal decomposition operation is approximately reversible, such that the entire process can be reversed.

Given that high-dimensional probability distributions from k-space data have complex multiscale properties. A key idea is that they can be simplified by factorizing them as a product for conditional probabilities of normalized wavelet coefficients via the discrete wavelet transform (DWT). These conditional probabilities are more comparable to Gaussian white noise than the original data distribution, and can thus be sampled more efficiently. Formally, a wavelet orthogonal transform decomposes original k-space data into low-frequency (LL) and high-frequency (LH, HL, HH) components for diffusion process in the subspace, which can be expressed as the following equation:

$$\{k^{LL}, k^{LH}, k^{HL}, k^{HH}\} = Wk, \quad (9)$$

$$\langle k^i, k^j \rangle = 0, \ i \neq j \in \{LL, LH, HL, HH\}, \quad (10)$$

where $W$ denotes the DWT and $k^{LL}$ indicates the low-frequency component encapsulates the principal features and structures, after the transform while $k^{LH}, k^{HL}, k^{HH}$ correspond to the high-frequency counterparts encapsulate detailed high-frequency elements in vertical, horizontal, and diagonal orientations, respectively. $\langle \cdot, \cdot \rangle$ denotes the inner product, which points out that the wavelet components are mutually orthogonal. Therefore, these orthogonal wavelet components satisfy the conditions for diffusion within the subspace. To simplify discussion in the following sections, we introduce a shorthand notation $\mathbb{K}$ encompassing all these frequency components in the wavelet-domain, i.e., $\mathbb{K} = \{k^{LL}, k^{LH}, k^{HL}, k^{HH}\}$.

### A. Score Matching in Subspace Diffusion Model

As previously discussed, the k-space data is first diffused in the full-space according to Eq. (3). To generate samples, we need to learn the function $\nabla_k \log p_t(k)$ as usual. However, the analytical form of $\nabla_k \log p_t(k)$ is generally intractable, and hence we learn a score model $s_{\theta_1}(k, \sigma_t)$ parameterized by the network to estimate its values:

$$\theta^* = \min_{\theta} \mathbb{E}_t\{\lambda_t \mathbb{E}_{k(0)} \mathbb{E}_{k(t)|k(0)}[||s_\theta(k(t), t) - \nabla_{k(t)} \log p_t(k(t)|k(0))||_2^2]\}. \quad (11)$$

Instead of conducting the score matching on the full-space, DWT is used to reduce the dimension of k-space data and then the diffusion process is restricted to the subspace in the interval $t \in (t_m, t_T)$. This means that wavelet components $\mathbb{K}$ as diffusion elements more accurately formulate the subspace. To learn the lower-dimensional diffusion process in subspace, we leverage the fact that the subspace components $\mathbb{K}$ of the k-space data diffuse under an SDE with the same $f(t)$ and $g(t)$ as the full-space, independent of the orthogonal components. Consider the case that we only use one proper subspace. Then the diffusion process in the subspace can be described as:

$$d\mathbb{K} = f(t)\mathbb{K}dt + g(t)du. \quad (12)$$

As a result, the perturbation kernels in the subspace have the same form as in the full-space, allowing us to train $s_{\theta_\mathbb{S}}(\mathbb{K}, t)$ to match the score $\nabla_\mathbb{K} \log p_t(\mathbb{K})$ via precisely the same procedure as in Eq. (11). These scores are related to the full-space scores $\nabla_k \log p_t(k)$ via $\mathbf{Q}$, but since $\mathbb{K} = \mathbf{Q}k$ for times $t > t_m$, we can directly work with data points $\mathbb{K}$ and score model $\nabla_\mathbb{K} \log p_t(\mathbb{K})$ with no loss of information. Thus, the model of subspace $s_{\theta_\mathbb{S}}(\mathbb{K}, t)$ can be trained with the following function:

$$\theta_\mathbb{S}^* = \min_{\theta_\mathbb{S}} \mathbb{E}_t\{\lambda_t \mathbb{E}_{\mathbb{K}(0)} \mathbb{E}_{\mathbb{K}(t)|\mathbb{K}(0)}[||s_{\theta_\mathbb{S}}(\mathbb{K}(t), t) - \nabla_{\mathbb{K}(t)} \log p_t(\mathbb{K}(t)|\mathbb{K}(0))||_2^2]\}. \quad (13)$$

Note that the score matching strategy remains consistent with the full-space, except that we treat the wavelet components as the original undiffused data. With Eq. (13), the training speed is significantly boosted due to the use of wavelet components.

### B. Iterative Sampling via Subspace Diffusion Prior

*Regularization Constraint:* To generate the samples, score models $s_\theta(k, t)$ and $s_{\theta_\mathbb{S}}(\mathbb{K}, t)$ in the corresponding interval $(t_m, t_T)$ are used to solve the reverse diffusion. Generically, the reverse diffusion process is equipped with the Predictor-Corrector sampler [14] in our implementation for its superior performance. Predictor is the SDE solver first gives an estimate of the reconstruction results at the next time step, while the Corrector is the score-based Markov Chain Monte Carlo approach refines the marginal distribution of the estimated results. Formally, the Predictor updates the estimate results in the full-space with the trained diffusion model $s_\theta(k, t)$ from the above section as:

$$k_{t-1} = k_t + (\sigma_t^2 - \sigma_{t-1}^2)s_\theta(k_t, t) + \sqrt{\sigma_t^2 - \sigma_{t-1}^2}Z, \quad (14)$$

where $\sigma_t$ represents a monotonically increasing function with respect to time $t \in (0, t_m)$. The variable $Z \sim N(0,1)$ follows a Gaussian distribution representing random noise. For each updated $k_{t-1}$, the following Corrector step is performed multiple times to refine it:

$$k_{t-1} = k_{t-1} + \varepsilon_{t-1}s_\theta(k_{t-1}, t) + \sqrt{2\varepsilon_{t-1}}Z, \quad (15)$$

where $\varepsilon_{t-1}$ is the step size at time $t-1$. Similar to the alternative sample strategy, data in the subspace can also be reconstructed by alternatively updating Predictor-Corrector samplers. Due to the reconstruction step is performed within the subspace, the original k-space data $k$ should first undergo orthogonal decomposition with the wavelet transform to produce $\mathbb{K}$. Hereafter, the Predictor step is conducted with the trained diffusion model $s_{\theta_\mathbb{S}}(\mathbb{K}, t)$ from the above section as follows:

$$\mathbb{K}_{t-1} = \mathbb{K}_t + (\sigma_t^2 - \sigma_{t-1}^2)s_{\theta_\mathbb{S}}(\mathbb{K}_t, t) + \sqrt{\sigma_t^2 - \sigma_{t-1}^2}Z. \quad (16)$$

Subsequently, the Corrector step is executed on $\mathbb{K}$ as follows:

$$\mathbb{K}_{t-1} = \mathbb{K}_{t-1} + \varepsilon_{t-1}s_{\theta_\mathbb{S}}(\mathbb{K}_{t-1}, t) + \sqrt{2\varepsilon_{t-1}}Z. \quad (17)$$

Predictor and Corrector samplers are alternatively performed for several times between the full-space and the subspace to reach convergence.

*Latent Consistency Module:* To preserve the accuracy of intricate appearance features in the reconstructed result, we further incorporate the data consistency between the reconstructed image and the measurement data within the Predictor-Corrector sampler. Considering the practical significance, the latent consistency module ensures that the intermediate solution remains in the feasible region of the data. However, since the diffused variables in the subspace undergo orthogonal decomposition based on wavelet components, the latent consistency module cannot directly extend over the subspace to constrain the sampling process. This suggests that an inverse DWT should be performed first, ensuring that the approximation error between the subspace updates and the measurement data remains below some tolerance threshold. In this case, the inverse wavelet transform can be defined as:

$$k_t^* = W^T \mathbb{K}_t, \tag{18}$$

where $k_t^*$ is the k-space data after subspace updating and $W^T$ denotes the inverse DWT. Following, we can enforce data consistency to guide the sampling process more legitimately in the subspace. The problem can be described as:

$$\min_k \{\|Ak - y\|_2^2 + \lambda \|k - k_t^*\|_2^2\}. \tag{19}$$

By integrating the latent consistency module into the unified diffusion framework, a mutual feedback loop can be formed with the regularization constraint.

Traditionally, the reconstruction problem has been conceptualized as a low-rank matrix completion problem. To achieve high-quality reconstruction results, we have incorporated traditional operator following network iteration, which serves to further restore the low-rank matrix. The object of traditional operator is a data matrix which is generated by the network. As the k-space data matrix transforms into Hankel matrix formulation, we can analyze the hard-threshold singular values of the data matrix. According to the low-rank property in Hankel matrix formulation, solving the low-rank constraint term turns to an optimization problem:

$$rank(L) = l, \ k = H^+(L), \tag{20}$$

where $H^+(\cdot)$ is the Hankel pseudo-inverse operator, $L$ is a data matrix with low-rank property after conducting hard-threshold singular values operation, and $l$ is the rank of the data matrix. Such a cooperative mechanism between the diffusion prior and the low-rank block could facilitate to address the challenges of overfitting of model training due to data redundancy.

## IV. EXPERIMENTS

### A. Experimental Setup

*Datasets:* In the main experiments, especially the training of the score function, were performed with the brain dataset ***SIAT***, which are provided by Shenzhen Institute of Advanced Technology, the Chinese Academy of Sciences and informed consent was obtained from the imaging object in compliance with the IRB policy. The fully-sampled SIAT data are acquired on a 3.0T Siemens Trio Tim MRI scanner, consisting of 12-channel complex-valued MR images with $256 \times 256$, and combined into single-channel data by coil compression. A subset of 500 samples is selected to form the training dataset. Subsequently, the dataset is expanded to 4000 single-coil images through flip and rotation, thereby ensuring a more comprehensive training to enhance model performance.

Noted that our Sub-DM is a one-for-all model. This means that once Sub-DM is trained, it can be reused for a diverse range of datasets with different sequences during the testing phase. One dataset was the ***T1-GE Brain***, including 8-channel complex-valued obtained by 3.0T GE. As well as the ***T1-weighted Brain***, obtained with an 8-channel joint-only coil using 1.5T GE. Additionally, the ***fastMRI+*** dataset [39] also is used to test the reconstruction performance of the proposed method, verifying its generalization ability. For the ***fastMRI+*** public brain data, we crop the target image size to $256 \times 256$ for direct reconstruction using the model trained on the ***SIAT*** brain dataset, and obtained accurate reconstruction details as expected. Code is available at: *https://github.com/yqx7150/Sub-DM*.

*Implementation Details:* The main parameter settings and training procedure of VE-SDE follow the guidelines recommended by [12]. Specifically, we controlled the noise variance schedule in forward diffusion by setting $\sigma_{max}=378$ and $\sigma_{min}=0.01$ to obtain prior information more effectively during training. For the Adam optimizer, we typically set $\beta_1 = 0.9$ and $\beta_2 = 0.999$ to optimize the network. It is also worth noting that to ensure data correction following a single Langevin sampling, $N = 1000$ and $M = 1$ are applied for the high-quality generation during reconstruction. For other parameters, we empirically tune within their recommended ranges to achieve optimal performance. All training and testing experiments are conducted using 2 NVIDIA 2080TI GPUs, 12 GB. Furthermore, the parameter settings of the sampling process remain consistent when we validate the effectiveness of the proposed model on all datasets. To quantitatively evaluate the reconstruction quality of the various reconstruction algorithm, we use metrics such as Peak Signal-to-Noise Ratio (PSNR), Structural Similarity Index (SSIM), and Mean Squared Error (MSE). To ensure a fair comparison, the results of each method are adjusted aiming to illustrate the advantages of Sub-DM.

### B. Reconstruction Experiments

*Comparisons with State-of-the-arts:* To evaluate the effectiveness of our proposed method, we compare its performance with other methods, including traditional methods SAKE [40] and ESPIRIT [41], as well as deep learning methods EBMRec [42] and HKGM [23] on two different datasets. For under-sampled MRI reconstruction under ×8, ×10, and ×12 acceleration factors across Poisson, Radial, Random, and Uniform sampling patterns, the PSNR, SSIM, and MSE values from the brain dataset are summarized in Table I. The quantitative results demonstrate that Sub-DM outperforms other reconstruction methods. Obviously, training within a subspace framework effectively learns profound information, which in turn facilitates the diffusion of effective features and enhances the robustness of the model. Even under higher acceleration factors, Sub-DM is capable of reconstructing more realistic details.

It is evident that the visual effects of Sub-DM are consistent with the quantitative results, compared to other methods under same acceleration factors in Fig. 4. Specifically, SAKE and ESPIRiT reconstructions exhibit noise, aliasing, and blurriness, resulting in degraded image quality and loss of fine structures. In contrast, EBMRec and HKGM, as the deep learning techniques, demonstrate substantial improvements in recovering the prominent structures and edges. Nevertheless, upon close examination, it becomes apparent that EBMRec inaccurately reconstruct structures, failing to capture the fine details, even suffers from heavy noise and artifacts under high acceleration factors. While the HKGM approach eliminates some artifacts, it results in a loss of high-frequency details, obtained unsatisfactory effects for noise suppression and structural detail preservation under high acceleration factors. Remarkably, Sub-DM achieves visual results with the most texture details and the least amount of noise, preserving the most realistic high-frequency details and effectively suppressing artifacts. Meanwhile, the enlarged views and error maps of the reconstruction further underscore the superiority of the proposed method over others.

TABLE I
PSNR, SSIM, AND MSE (*10$^{-4}$) COMPARISON WITH STATE-OF-THE-ART METHODS UNDER POISSON, RADIAL, 2D RANDOM, AND UNIFORM SAMPLING PATTERNS WITH VARYING ACCELERATION FACTORS.

| *T1-GE Brain* | AF | SAKE [40] | ESPIRIT [41] | EBMRec [42] | HKGM [23] | Sub-DM |
|---|---|---|---|---|---|---|
| Poisson | R=8 | 37.20/0.9149/1.493 | 33.49/0.8976/4.473 | 30.75/0.8097/8.403 | 38.45/0.9485/1.430 | **42.11/0.9544/0.614** |
|  | R=12 | 34.25/0.8856/3.845 | 29.88/0.8652/10.288 | 28.63/0.7731/13.696 | 34.73/0.9093/3.366 | **40.71/0.9341/0.849** |
| Radial | R=10 | 31.76/0.8939/8.782 | 28.41/0.8043/14.407 | 29.40/0.7839/11.478 | 32.66/0.9024/5.416 | **38.67/0.9464/1.357** |
|  | R=12 | 29.66/0.8646/9.587 | 27.37/0.7685/18.302 | 27.39/0.7195/18.210 | 32.52/0.8849/5.596 | **36.33/0.9316/2.327** |
| *T1-weighted Brain* | AF | SAKE [40] | ESPIRIT [41] | EBMRec [42] | HKGM [23] | Sub-DM |
| Random | R=10 | 27.52/0.6434/61.826 | 29.15/0.7470/12.165 | 27.75/0.5820/16.778 | 28.21/0.5816/15.096 | **37.34/0.8396/1.845** |
|  | R=12 | 26.62/0.6288/73.287 | 28.56/0.7236/13.927 | 26.48/0.4930/22.469 | 26.66/0.5441/21.558 | **36.57/0.8407/2.201** |
| Uniform | R=8 | 29.63/0.7648/13.215 | 31.81/0.7511/6.595 | 28.62/0.5899/13.755 | 31.47/0.8456/7.124 | **36.60/0.8941/2.188** |
|  | R=10 | 28.83/0.7460/17.289 | 29.24/0.7475/11.906 | 28.18/0.5778/15.216 | 30.92/0.8375/8.098 | **36.33/0.8894/2.327** |

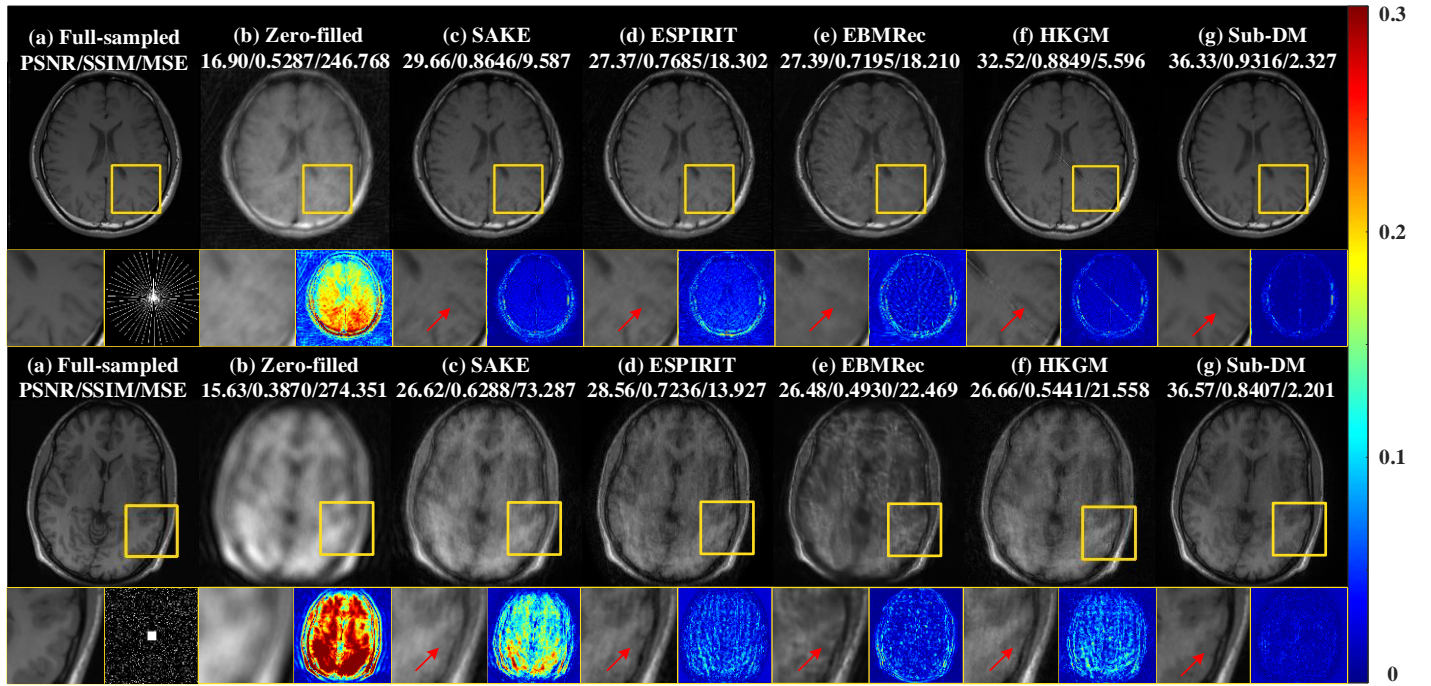

**Fig. 4.** Reconstruction results of *T1-GE Brain* and *T1-weighted Brain* under Radial (first two rows) and Random (last two rows) sampling at R=12. The first row shows (a) full-sampled (b) zero-filled, reconstruction by (c) SAKE, (d) ESPIRIT, (e) EBMRec, (f) HKGM, and (g) Sub-DM. The second row shows the enlarged views indicated by the selected fine detailed regions, and the error map of the reconstruction.

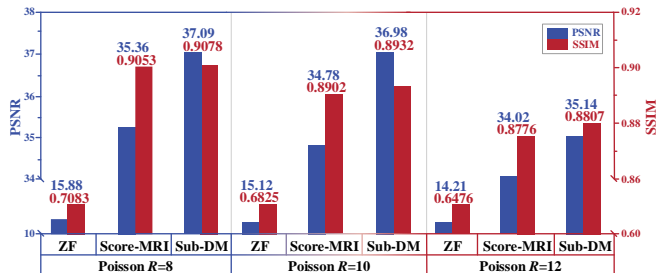

**Fig. 5.** The histograms of PSNR and SSLM values at different acceleration factors (R=8, 10, 12) using zero-filled (ZF), Score-MRI, and Sub-DM methods under Poisson sampling.

*Advantages of Low-dimensional Subspace:* To highlight the advantages of subspace over the full-space, a quantitative comparison was performed between Sub-DM and the typical full-space method Score-MRI [15] in Fig. 5. It clearly demonstrates that Sub-DM achieves the best performance relative to the Score-MRI across all acceleration factors. Visual results are provided under the Poisson sampling pattern at acceleration factor of R=10 in Fig. 6. As seen from the region of interest (ROI) marked with yellow squares, significant lack of detailed textures can be clearly observed in the result of Score-MRI when compared with the Sub-DM. These differences arise from the network's requirement to learn score functions in high-dimensional space, which hampers effective assessment of the score models and leads to the loss of high-frequency information. In contrast, the proposed method migrates the diffusion process to the subspace, which then learn to match a lower-dimensional score function may lead to refined training coverage and further improved performance.

*Effects of Distinct Subspace:* A comparative experiment was also conducted with the distinct subspace method HFS-SDE [24], and quantitative metrics under uniform sampling are provided in Fig. 7. In comparison, Sub-DM achieved significantly higher metrics under all acceleration factors. Especially, the PSNR value of the reconstructed image by Sub-DM increases by 1.42 dB and SSIM remains at 0.9 compared to HFS-SDE at the acceleration factor of R=8. Fig. 8 precisely depicts the visual effects of Sub-DM and competing method. It can be seen that Sub-DM has much lower errors with finer detailed tissue structures and more edge information, indicating Sub-DM effectively captures more prior information. Specifically, unlike the HFS-SDE diffused in a high-dimensional subspace, Sub-DM adopts a subspace dimensionality reduction strategy to separate and learn different regions or details in the image, and promotes the generation of more realistic MRI details during reconstruction by a low-dimensional score network.

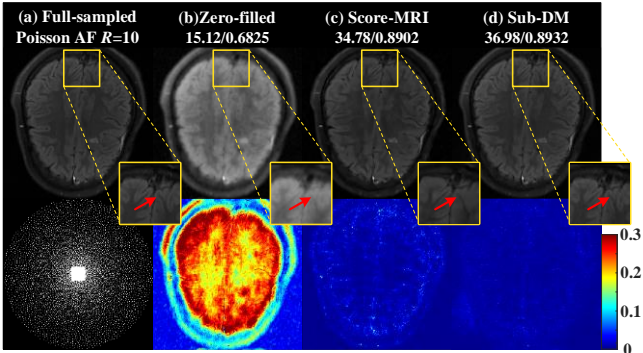

**Fig. 6.** Reconstruction results of Score-MRI and Sub-DM under Poisson sampling at $R$=10. The values on the top are PSNR/SSIM values. Second row illustrates the error map. The windows in the middle are detail enlarged areas.

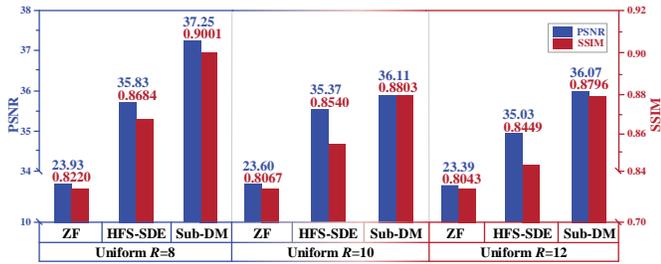

**Fig. 7.** PSNR and SSIM values across ×8, ×10, and ×12 acceleration factors, obtained by ZF, HFS-SDE, and Sub-DM methods under Uniform sampling.

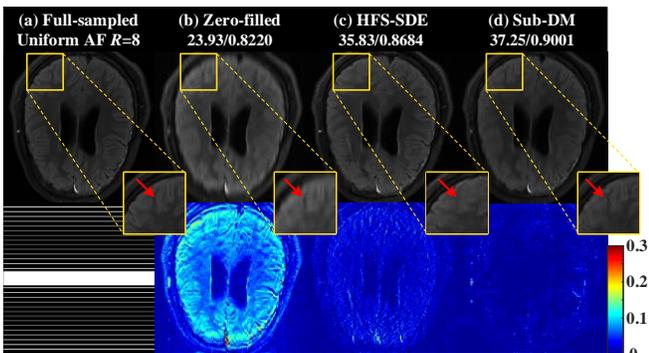

**Fig. 8.** The comparison of reconstruction results for HFS-SDE and Sub-DM under Uniform mask at $R$=8. Second row shows the error views. The extracted ROI further demonstrates the advantages of our proposed method.

*Necessity of Orthogonal Decomposition:* Experiments are also conducted using WKGM [30] to compare different k-space optimization methods. As evident from the quantitative results in Fig. 9, Sub-DM consistently outperforms WKGM across all cases. Furthermore, as the acceleration factor is increased, WKGM exhibits a corresponding decrease in reconstruction performance, which is consistent with the observations in Fig.

10. Concretely, some noise still remains in the results of WKGM, hindering the full preservation of structural details. In contrast, Sub-DM can easily identify over-smoothing and distortion, while reconstructing accurate texture details with less noise. Such results show that although the k-space weighting technique effectively optimizes the network reconstruction, it does not capture the difference between high and low frequencies. While Sub-DM decomposes the k-space into low-dimensional orthogonal components with DWT, effectively learns the distinctive characteristics of both high and low-frequency priors, reducing the training computational cost and achieving high-quality reconstructions.

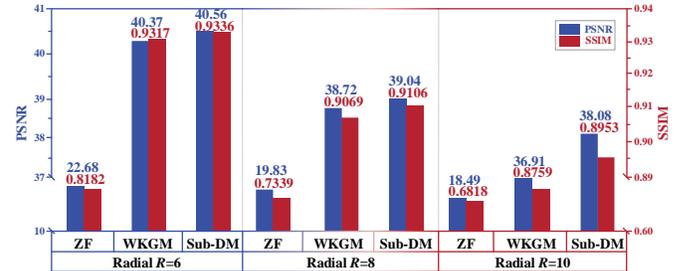

**Fig. 9.** The histograms of PSNR and SSIM values for ZF, WKGM, and Sub-DM methods under Radial mask at acceleration factors of $R$=6, 8, 10.

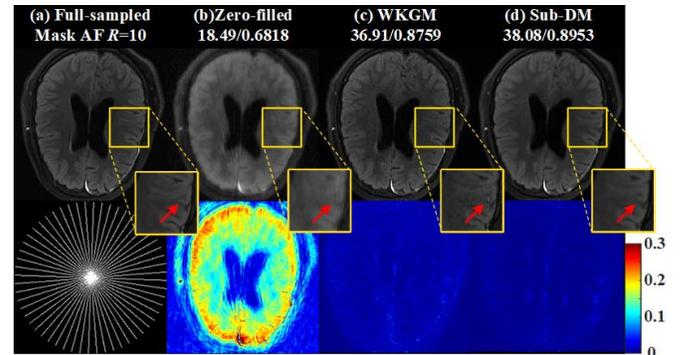

**Fig. 10.** Reconstruction results using WKGM and Sub-DM at $R$=10 of Radial mask. From left to right: full-sampled, zero-filled, WKGM, and our Sub-DM. The second row displays error views. The yellow box represents the ROI, with an amplified view.

### C. Convergence Analysis and Computational Cost

*Convergence Analysis:* In this subsection, comparison with DM-based MRI reconstruction methods on *fastMRI+* dataset was conducted. The convergence plot in Fig. 1 highlights the performance of Sub-DM in comparison to other MRI reconstruction methods, demonstrating its clear superiority. Sub-DM not only achieves the highest PSNR, but also has significantly fewer iteration steps, around 50, surpassing other methods in both speed and image quality. In contrast, the other models, including HKGM, WKGM, and HFS-SDE, require more iterations to reach lower PSNR levels. CSGM-MRI and Score-MRI are the slowest to converge, needing over 1500 and 2000 iterations, respectively, and still fall short in terms of PSNR. The specific computational cost is shown in Table II. Consequently, it can be concluded that migrating the diffusion process to the subspace effectively promotes the capture of high-frequency information and accelerates convergence. Moreover, the correlation between different orthogonal components increases the diversity of the prior distribution, facilitating the learning of a lower-dimensional score function, which results in more refined training coverage and improved model performance.

*Practical Considerations on Iteration Time:* Given the evident constraint of image reconstruction for using diffusion models is the sampling time, we further explored the diffusion time in full-space and subspace method under practical experimental conditions. According to the results in Table II, it can be seen that the algorithm Sub-DM, where the diffusion process is migrated in the subspace, rapidly achieves convergence with relatively low time consumption. Meanwhile, limited by the inherent characteristics of the diffusion model in the entire space, score-MRI and WKGM forward process occurs in the entire environment space of the data distribution, and its high dimensionality further increases the computational cost and increases sampling time. The integration of subspace and orthogonal decomposition strategies undoubtedly reduces runtime, and maximizing their synergy further balances reconstruction performance with iteration time.

TABLE II
COMPUTATIONAL COST (UNIT: SECOND) OF DIFFERENT METHODS.

| Method | Score-MRI | WKGM | Sub-DM |
|---|---|---|---|
| Total-time (s) | 26120 | 6718.4 | **573.72** |
| Iter-time (s) | 13.06 | 13.6 | **13.66** |
| Iter-numbers | 2000 | 494 | **42** |

## V. Discussion

The previous section has been proved that the strategy of diffusing k-space data in the subspace via orthogonal decomposition can significantly enhance both the reconstruction quality and convergence speed, while also reducing the time required for sampling without compromising the reconstruction quality. However, certain aspects of our proposed model still require further discussion and improvement.

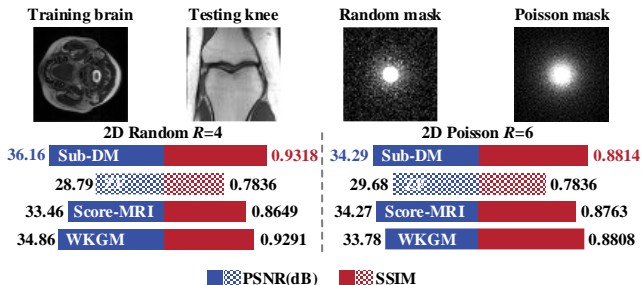

Fig. 11. Experimental results of different methods in terms of PSNR and SSIM in out-of-distribution reconstruction tasks with 2D Random and Poisson mask.

*Out-of-Distribution Performance:* Assessing the practical generalizability of deep learning techniques is of considerable importance in MRI reconstruction. This focus can lead to degraded performance when training and testing data distributions diverge and impedes efficient feature extraction for out-of-distribution samples. However, the subspace approach that promotes the diffusion of informative features demonstrates resilience and flexibility across diverse reconstruction challenges, independent of superficial data patterns. To validate this hypothesis, the experimental design was trained on the SIAT brain dataset, followed by reconstruction on the *fastMRI* knee dataset. The results, as presented in Fig. 11, demonstrate that the proposed method has good generalization capabilities, accurately reconstructing images that are out-of-distribution.

*Low-rank Optimization:* It is also significant to analyze the influence of the optimization module shown in Fig. 3 on the reconstruction results. Low-rank matrix completion strategy leverages the redundancy within image data to mitigate noise and augment image quality, proving particularly effective in scenarios where data is incomplete or corrupted. Therefore, we performed a quantitative evaluation of the low-rank optimization module, with the corresponding metrics results presented in Table III. Obviously, the metrics listed in Table III illustrate the contribution of this optimized strategy to overall performance enhancement, and the impact of the optimization components on the reconstruction process is found to be constrained. At the same time, even in the absence of the low-rank module, Sub-DM, with its robust generative capabilities, demonstrates excellent performance under high acceleration factors. Hence, combining the two can complement each other, leading to an overall improvement in reconstruction performance.

TABLE III
COMPARISON OF PSNR AND SSIM UNDER RANDOM SAMPLING MODES WITH DIFFERENT ACCELERATION FACTORS.

| *T1-GE Brain* | Random | | |
|---|---|---|---|
| **Low-rank** | **×4** | **×5** | **×6** |
| × | 43.40/0.9752 | 41.95/0.9679 | 40.99/0.9629 |
| √ | **46.26/0.9793** | **44.55/0.9718** | **43.69/0.9644** |

## VI. Conclusion

In summary, the primary focus of the current study was a subspace diffusion model for MRI reconstruction that combined orthogonal decomposition to simplify complex distributions, aiming to enhance reconstruction accuracy while increasing reconstruction rate. Specifically, we migrated the diffusion process to a subspace and decomposes the k-space data into multiple orthogonal subcomponents. This method facilitated the learning of effective information within complex distributions, addressing the inherent high-dimensional extrapolation challenges and reducing the computational cost associated with diffusion models. Extensive theoretical analysis and rigorous experiments have shown that Sub-DM is not only competitive in in-distribution reconstruction tasks, but also achieves better reconstruction in OOD. Therefore, it remains important future work to explore the potential benefits of adopting subspace operations in MRI reconstruction.